\documentclass[10pt,amssymb,twocolumn,notitlepage]{article}

\usepackage{epigraph}
\usepackage[affil-it]{authblk}
\usepackage{dcolumn}
\usepackage[stable]{footmisc}
\usepackage{blkarray}
\usepackage{bm}
\usepackage[mathscr]{euscript}
\usepackage[font=scriptsize]{caption}
\usepackage{subcaption,graphicx}
\usepackage{amsmath}
\usepackage[table]{xcolor}
\usepackage[margin=1.7cm]{geometry}
\usepackage{tcolorbox}
\usepackage{hyperref}
\hypersetup{
    colorlinks=true,
    linkcolor=blue,
    filecolor=magenta,      
    urlcolor=blue,
}
\usepackage{gensymb}
\usepackage[font=footnotesize]{caption}
\usepackage{hyperref}
\usepackage{amssymb}
\usepackage{makecell}
\usepackage{algorithm,lipsum,xcolor,caption}
\usepackage{enumitem}
\usepackage{esint}
\usepackage{fancyvrb}
\usepackage{xcolor}
\usepackage{wasysym}
\usepackage{arydshln}% http://ctan.org/pkg/arydshln

\definecolor{newblue}{rgb}{0.0, 0.28, 0.67}
\definecolor{newgreen}{rgb}{0.13, 0.55, 0.13}
\definecolor{newred}{rgb}{0.87, 0.72, 0.53}

\definecolor{newblue}{rgb}{0.0, 0.28, 0.67}
\definecolor{newgreen}{rgb}{0.13, 0.55, 0.13}
\definecolor{newred}{rgb}{0.87, 0.72, 0.53}

\title{Complex Networks of Functions}
\author{Luciano da Fontoura Costa \\ \emph{luciano@ifsc.usp.br}}
\affil{S\~ao Carlos Institute of Physics -- DFCM/USP \\
Av. Trab. S\~ao Carlense, 400\\
S\~ao Carlos - SP, 13566-590, Brazil } 
\date{20th Jan. 2021}

\begin{document}

\twocolumn[
\begin{@twocolumnfalse}
    \maketitle
    \begin{abstract}
    Functions correspond to one of the key concepts in mathematics and science, allowing the
    representation and modeling of several types of signals and systems.  The present work 
    develops an approach for characterizing the coverage and interrelationship between 
    discrete signals that can be fitted by a set of reference functions, allowing the definition 
    of transition networks between the considered discrete signals.  While the adjacency
    between discrete signals is defined in terms of respective Euclidean distances, the
    property of being adjustable by the reference functions provides an additional constraint
    leading to a surprisingly diversity of transition networks topologies.    First, we motivate the 
    possibility to define transitions between parametric continuous functions, a concept that 
    is subsequently extended to discrete functions and signals.   Given that the set of all possible 
    discrete signals in a bound region corresponds to a finite number of cases, it becomes 
    feasible to verify the adherence of each of these signals with respect to a reference set of 
    functions.  Then, by taking into account also the Euclidean proximity between those discrete 
    signals found to be adjustable, it becomes possible to obtain a respective transition network 
    that can be not only used to study the properties and interrelationships of the involved discrete 
    signals as underlain by the reference functions, but which also provide an interesting complex 
    network theoretical model on itself, presenting a surprising diversity of topological features, 
    including modular organization coexisting with more uniform portions, tails and handles, as 
    well as hubs. Examples of the proposed concepts and methodologies are provided respectively
    with respect to three case examples involving power, sinusoidal and polynomial functions.
    \end{abstract}
\end{@twocolumnfalse} \bigskip
]

\setlength{\epigraphwidth}{.49\textwidth}
\epigraph{`\emph{Qualcosa corre tra loro, uno scambiarsi di sguardi come linee che collegano 
una figura all\textsc{\char13}altra e disegnano frecce, stelle, triangoli ...}'}{Italo Calvino, Le citt\`a invisibili.}

\section{Introduction}

Functions are to mathematics as sentences are to linguistics, constituting 
basic resources for develping more complete mathematical systems and models.  
The importance of functions is reflected in their widespread applications 
not only to the physical sciences, but to virtually every scientific field.

Traditionally, the mathematical study of functions and their properties has been
approached in continuous vector spaces, involving infinite instances of a given type
of function.  While this constitutes an effective and important approach, most of the 
signals in practical applications have discrete nature, being represented as discrete 
signals or vectors. This follows as a consequence of the sampling of physical signals by
using acquisition systems that inherently implies the signals to be quantized along their
domain and magnitude.

Though discrete functions are systematically studied in areas as digital signal
processing (e.g.~\cite{oppenheim:2009,Parr:2013}), emphasis is often placed on aspects of
quantization errors and representations in the frequency domain, employing the
Fourier series or transform (e.g.~\cite{brigham:1988}).  However, relatively lesser attention is 
typically focused on the relationship between the discrete signals, or on how they can be 
approximated by specific functions.   Though the latter subject constitutes one of the
main motivation of the areas of numerical methods (e.g.~\cite{recipes}) and numerical 
analysis (e.g.~\cite{Burden:2015}), this subject is typically approached from the perspective of function 
approximation, not often addressing the interrelationship between functions.  

The present work develops an approach aimed at characterizing not only which
discrete signals in a discrete region $\Omega \subset \Re^2$ can be 
adjusted by a given set of reference functions $g_i(x)$, $i = 1, 2, \ldots, N$, but also
how such adjustable discrete signals interrelate one another in the sense of
being similar, or adjacent.   The consideration of a pre-specified set of function
types happens frequently in science, especially when fitting data or studying 
dynamic systems.  In particular, the solution of linear systems of differential equations
is often approached in terms of linear combinations of a set of eigenfunctions
(e.g.~\cite{NagleSaff}), 
which could also be taken as the reference functions considered in this work.

The concepts and methods developed in the present article are interesting not only
theoretically while studying how distinct types of functions are related, but also from
several application perspectives, such as characterizing specific discrete spaces,
discrete signal approximation, morphing of functions (i.e.~transforming a function
into another through incremental changes), controlling systems underlain by specific
types of function, among many other possibilities.  In a sense, functions can be
approached as a way to constrain, in specific respective manners,
the adjacency between continuous signals in a given region or space.  For instance,
the function sine restricts all possible continuous signals in a given space.

In addition to the relevance of the described developments respectively to the 
aforementioned mathematical aspects, they also provide several contributions
to the area of network science (e.g.~\cite{netwsci}).  Indeed, as it will be seen along this work, the
transition networks derived from discrete signal spaces with respect to sets of
reference functions are characterized by a noticeably rich topological structure
that can involve modularity, hubs, symmetries, handles and tails~\cite{tails_handles}, as well
as coexistence of regular and modular subgraphs.  As such, these networks
provide valuable resources not only regarding the characterization of complex
networks, but also constitute a model or benchmark that can be used as reference in studies
aimed at investigating the classification and robustness of networks, as well
as investigations addressing the particularly challenging relationship between
network topology and possible implemented dynamics.

In order to obtain the means for quantifying how discrete signals in a region
can be approximated by reference functions, and how these signals interrelate
one another, we develop several respective concepts and methods.   More specifically,
after defining the problem in a more formal manner,  we proceed by suggesting
how to define a system of adjacencies between continuous functions, in terms
of the identification of respective transition points.  These concepts are then
transferred to discrete signals, allowing the proposal of indices for quantifying
the coverage of the discrete signals by adopted reference functions.  Subsequently,
we adapt the concept of adjacency between functions to discrete signals, 
allowing the derivation of a methodology for obtaining transition networks expressing
how the discrete signals in a region can be transformed one another while being
approximated as instances of the reference functions.  Several studies involving
the obtained transition networks are then described, including the identification of
shortest paths between two adjustable discrete signals, random walks related to the
unfolding of dynamics on the network, as well as the possibility of identifying 
discrete signals that are more central regarding the interrelationships represented
in the transition networks.

The developed concepts and methods are then illustrated with respect to three main
case examples involving (i) four power functions; (ii) a single complete polynomial of
forth order; and (iii) two sets of  hybrid reference  functions involving combinations of  
power functions  and sinusoidals.  Several remarkable results are identified and
discussed.

\section{Defining the Problem}

\begin{figure}[h!]  
\begin{center}
    \includegraphics[width=0.7\linewidth]{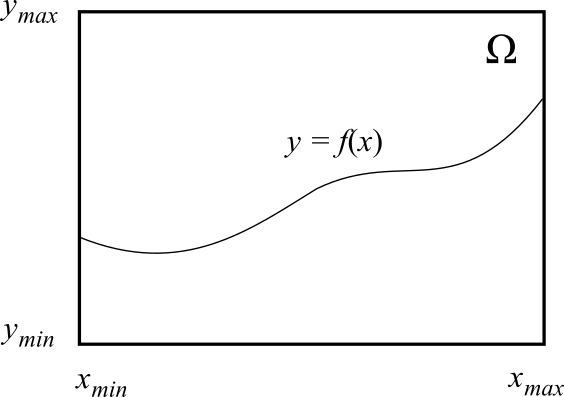}  \\  \vspace{0.1cm}
    \caption{A region $\Omega \subset Re^2$ delimited as $x_{min} \leq x \leq x_{max}$ 
    and $y_{min} \leq y \leq y_{max}$, and an example of a function $y = f(x)$
    completely comprised in this region.}
    \label{fig:cont}
    \end{center}
\end{figure}

Consider the region $\Omega \subset \Re^2$ in Figure~\ref{fig:cont}, which corresponds to
the Cartesian product of the intervals $x_{min} \leq x \leq x_{max}$ and 
$y_{min} \leq y \leq y_{max}$.

Let $y = f(x)$ correspond to a generic \emph{signal}, which can be associated 
to a function, completely bound in $\Omega$,
in the sense of having all its points comprised within $\Omega$.  No requirement,
such as continuity or smoothness, are whatsoever imposed on these functions.  

In addition, consider the \emph{difference} between two generic functions
 $y = f(x)$ and $y = h(x)$, both comprised in $\Omega$, as corresponding to
 the following root mean square distance (or error):
 \begin{equation} 
   \delta(f,h) = \sqrt{ \frac{1}{x_{max}-x_{min}} \int_{x_{min}}^{x_{max}}  [f(x) - h(x) ]^2 dx }
 \end{equation}
 
A possible manner to quantify the \emph{similarity} between $f(x)$ and $h(x)$ is as
\begin{equation}  \label{eq:simil}
   \sigma(f,h) = e^{-\alpha \, \delta(f,h)}
\end{equation}
 
for some chosen value of $\alpha$.
 
Let $y = g_i(x)$, $i = 1, 2, \ldots, N$ be a finite set of specific functions \emph{types} 
taken as a reference for our analysis.  For instance, we could have $g_1(x) = a_1 x + a_0$,   
$g_2(x) = a_1 x^2 + a_0$, and $g_3(x) = a_1 x^3 + a_0$ and $g_4(x) = a_1 x^4 + a_0$,
with $a_0, a_1, a_2, a_3, a_4 \subset \Re$.

An interesting question regards the identification, among all the possible signals
$y = \tilde{f}(x)$ in $\Omega$, of which of these signals can be expressed as $g_i(x)$ for 
$i = 1, 2, \ldots, N$, by yielding zero difference or unit similarity between $\tilde{f}(x)$ and
$g_i(x)$.    For each of the reference functions $g_i(x)$, we obtain a respective
set $S_i$ containing all functions $\tilde{f}(x)$ that can be exactly expressed in terms
of $g_i(x)$.

It is also interesting to allow for some tolerance by taking these two functions to
be related provided:
\begin{equation}
   \delta(\tilde{f},g_i)  \leq \tau_d
\end{equation}

or, considering their similarity, as:
\begin{equation}
   \sigma(\tilde{f},g_i)  \geq \tau_s
\end{equation}

with:
\begin{equation}  
   \tau_s = e^{-\alpha \, \tau_d}
\end{equation}

The identification of the sets $S_i$ can provide interesting insights regarding the
relative density of each type of the reference functions in the specified region
$\Omega$, paving the way to the identification of reference functions with more
general fitting capability as well as the interrelationship between these functions, 
in the sense of their proximity.

It should be observed that the obtained $S_i$ will also depend on the specific
size (or even shape) of $\Omega$, as a consequence of the requirement of all
functions to be completely bound in that region.  The alternative approach of
allowing the clipping of functions can also be considered, but this is not developed
in the present work.

In this work, we focus on \emph{discrete signals}, which are typically handled in scientific
applications and technology.  These signals are sampled along their domain and
quantized in their magnitude (see Section~\ref{sec:discrete}).
In order to identify the adjacency between discrete signals, given an $\Omega$ and a set of 
reference functions $g_i(x)$, first we identify (by using linear least squares) the discrete 
functions that can be approximated, within a tolerance, by the reference functions, therefore
defining the sets $S_i$, and then 
link these functions by considering their pairwise Euclidean distance.  
The thus obtained network $\Gamma$ or network can be verified to be 
undirected  and to contain a total of nodes equal to the sum of the cardinality of the obtained 
sets $S_i$, $i = 1, 2, \ldots, N$.  

In addition, each of the nodes becomes intrinsically associated
to the respective reference functions that were found to provide a good respective 
approximation.  In case a discrete function $\vec{f}$ is found to be adjusted by two or more of the
reference functions, only that corresponding to the best fitting may be associated to $\vec{f}$,
therefore avoiding replicated labeling.  The reference function thus associated to each node
of $\Gamma$ is henceforth called the node \emph{type}.

The transition network $\Gamma$ provides a systematic representation of
the relationships between the discrete functions in $\Omega$ that can be reasonably 
approximated by the reference functions.  Several concepts and
methods from the area of network science (e.g.~\cite{netwsci,surv_meas}) can then 
be applied in order to characterize the topological properties of the obtained network.  
For instance, the average
degree of a node can provide an interesting indication about how that function
can be transformed (or `morphed'), by a minimal perturbation, into other functions in $\Omega$.

The definition of a system of adjacencies between the functions of $\Omega$ as proposed
above also paves the way for performing respective random walks (e.g.~\cite{Justin}).  Starting at a given node,
adjacent nodes are subsequently visited according to a given criterion (e.g.~uniform
probability), therefore defining sequences of incremental transformations of the original
function.  These trajectories of functions can provide insights about how a function can be
progressively transformed into another (morphing), to define minimal distances between
any of the adjustable functions in $\Omega$ or, when associated to energy landscapes,
to investigate the properties of respectively associated dynamical systems (e.g.~\cite{Ogata}), 
including possible  oscillations (cycles) and chaotic behavior.

\section{Continuous Function Adjacency} \label{sec:adjacency}

A mathematical function often involves parameters, corresponding to values determining
its respective instantiation.  For instance, the function:
\begin{equation}
  g(x) = a_1 x + a_0
\end{equation}

corresponds to a straight line function whose inclination and translation is specified
by the parameters $a_0$ and $a_1$, respectively.

Given two generic functions in the region $\Omega$, a particularly interesting question
is whether one of them can be made identical to the other, which will be henceforth
be expressed as these functions being mutually \emph{adjacent}, in the sense of
providing an interface between these two functions, which can be therefore
transitioned.  More specifically, let the
two following functions $g_i(x)$ and $g_j(x)$, with respective parameters 
$a^i_0, a^i_1, \ldots, a^i_{N_i}$ and $a^j_0, a^j_1, \ldots, a^j_{N_j}$:
\begin{eqnarray}
  g_i(x; a^i_0, a^i_1, \ldots, a^i_{N_i})  \nonumber \\
  g_j(x; a^j_0, a^j_1, \ldots, a^j_{N_j})  \nonumber 
\end{eqnarray}

It should be kept in mind that, throughout this work, the superscript
value $j$ in the terms $a^j_0$ corresponds to an 
index associated to the respective reference function, not corresponding to the $j$-power of
$a$.

The functions $g_i()$ and $g_j()$ can be said to be \emph{adjacent} provided it is possible to find 
respective configurations of  parameters $ \tilde{a}^i_0, \tilde{a}^i_1, \ldots, \tilde{a}^i_{N_i}$ 
and $\tilde{a}^j_0, \tilde{a}^j_1, \ldots, \tilde{a}^j_{N_j}$ so that:
\begin{equation}
    g_i(x; \tilde{a}^i_0, \tilde{a}^i_1, \ldots, \tilde{a}^i_{N_i})  =  g_j(x; \tilde{a}^j_0, \tilde{a}^j_1, \ldots, \tilde{a}^j_{N_j})  
\end{equation}

for every value of $x$ in $\Omega$.

The set of parameters $\tilde{a}^i_0, \tilde{a}^i_1, \ldots, \tilde{a}^i_{N_i}$ 
and $\tilde{a}^j_0, \tilde{a}^j_1, \ldots, \tilde{a}^j_{N_j}$ are henceforth understood to represent
a \emph{transition point} in the parameter space $[\tilde{a}^i_0, \tilde{a}^i_1, \ldots, \tilde{a}^i_{N_i}, \tilde{a}^j_0, \tilde{a}^j_1, \ldots, \tilde{a}^j_{N_j}]$, namely:
\begin{equation}
  P_{g_i \leftrightarrow g_j}:   [\tilde{a}^i_0, \tilde{a}^i_1, \ldots, \tilde{a}^i_{N_i}, \tilde{a}^j_0, \tilde{a}^j_1, \ldots, \tilde{a}^j_{N_j}]
\end{equation}

Observe that each transition point defines a respective
instantiation of both involved functions, therefore also corresponding to a specific 
instantiated function in $\Omega$.

As an example, let's consider the following four parametric power functions:
\begin{eqnarray} \label{eq:four_ref}
    g_1(x) = a^1_1 x + a^1_0 \nonumber \\
    g_2(x) = a^2_1 x^2 + a^2_0 \nonumber \\
    g_3(x) = a^3_1 x^3  + a^3_0 \nonumber \\
    g_4(x) = a^4_1 x^4  + a^4_0  
\end{eqnarray}

with $a^1_0, a^2_0, a^3_0, a^4_0, a^1_1, a^2_2, a^3_3, a^4_4 \subset \Re$.

\begin{figure}[h!]  
\begin{center}
    \includegraphics[width=0.45\linewidth]{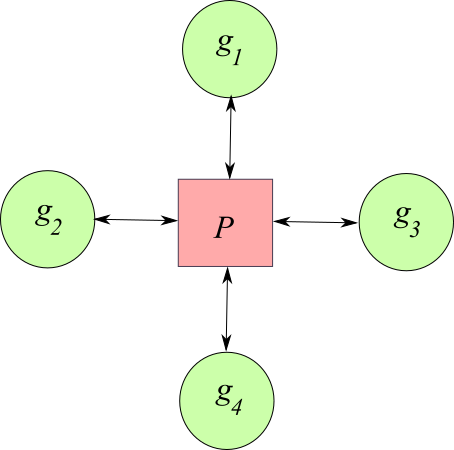}  \\  \vspace{0.1cm}
    \caption{The four reference functions in Eq.~\ref{eq:four_ref} share the transition point
    $P$ given as $a_0 \in [y_{min},y_{max}]$ with  $\tilde{a}^1_1 = \tilde{a}^2_1 = \tilde{a}^3_1 = \tilde{a}^4_1 = 0$.
    For one of the reference functions $g_i$ to transition to another function $g_j$, it
    is necessary that $g_i$ be instantiated to the function corresponding to $P$ through
    a respective parameter configuration, from which it can then follow to $g_j$.  Observe that,
    though this diagram involves only five basic nodes (functions), there is actually an infinite number of
    respectively defined situations in $\Omega$ as a consequence of its continuous nature.}
    \label{fig:trans_net}
    \end{center}
\end{figure}

All pairwise combinations of these functions $g_i()$ and $g_j()$ have respective transition points 
corresponding to $\tilde{a}^i_1 = \tilde{a}^j_1 = 0$ for any values of 
$a^i_0$ and $a^j_0$ for which the functions remain completely comprised within $\Omega$.
Though the four reference functions above have an infinite number of pairwise transitions points, 
each of them defines a respective transition network $\Gamma$ as presented in Figure~\ref{fig:trans_net}.

It is interesting to observe that, in this particular example, each of the transition points corresponds
to the constant functions $g(x) = a_0 = a^1_0 = a^3_0 = a^4_0 $, which therefore acts as a 
quadruple transition point for each $a_0 \in [y_{min},y_{max}]$ with 
$\tilde{a}^1_1 = \tilde{a}^2_1 = \tilde{a}^3_1 = \tilde{a}^4_1 = 0$:
\begin{equation}
  P:   [\tilde{a}^1_0 = \tilde{a}^2_0= \tilde{a}^3_0 = \tilde{a}^4_0 = a_0, \tilde{a}^1_1 = \tilde{a}^2_1 = \tilde{a}^3_1 = \tilde{a}^4_1 = 0]
\end{equation}

Observe that other sets of reference functions
can present many other types of transition points, which can be of types other than the null function.
Actually, any shared term between two parametric functions potentially corresponds to a
transition point.

In addition to transitions between types of reference functions as developed above, it is also
possible to have transitions between incrementally different instances of a same type of function.
This can be achieved by adopting a tolerance $\tau$ regarding the similarity of two
instances of the same type of function $g(x)$, i.e.: 
\begin{eqnarray}
    \int_{x_{min}}^{x_{max}} [ g(x; a_0, a_1, \ldots, a_{N - 1})  - \nonumber \\
     - g(x; a_0 + \delta_0, a_1+ \delta_1, \ldots, a_{N_i}+ \delta_{N-1}) ]^2 dx \leq \tau  \nonumber
\end{eqnarray}

In this manner, it is possible to obtain long sequences of transitions between instances
of a same function as the respective parameters are incrementally variated ($\vec{\delta}$), 
typically giving rise
to handles and tails~\cite{tails_handles} in respectively obtained network representations.  

Given that the approach reported in this work is respective to discrete signals and functions
given a tolerance $\tau$, both types of function transitions identified in this section are
expected to be taken into account and incorporated into the respectively derived transition
networks.

\section{The Discrete Case}  \label{sec:discrete}
 
Though interesting in itself, the above described problem involves infinite and non-countable 
sets  $S_i$.  Though this could be approached by using specific mathematical resources,
in the present work we focus on regions $\Omega$ that are discrete in both $x$ and 
$y$, taken with respective resolutions:
\begin{eqnarray}
   \Delta x = \frac{x_{max} - x_{min}}{N_x - 1}   \nonumber \\
   \Delta y = \frac{y_{max} - y_{min}}{N_y - 1}
\end{eqnarray}

where $N_x$ and $N_y$ correspond to the number of discrete values taken for
representing $x$ and $y$, respectively.

\begin{figure}[h]  
\begin{center}
    \includegraphics[width=0.7\linewidth]{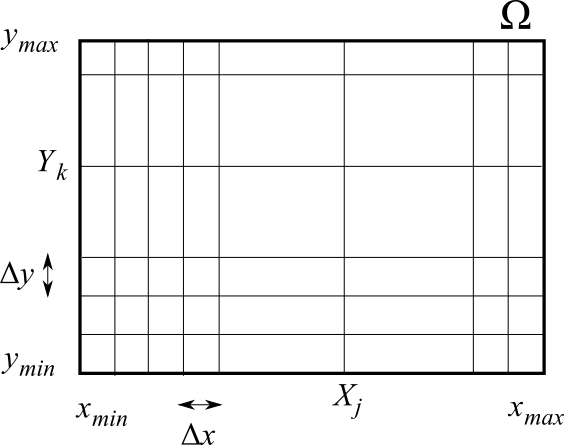}  \\  \vspace{0.1cm}
    \caption{A discretized region $\Omega \subset Re^2$, with $N_x$ values
    along the $x$-axis and $N_y$ values along the $y$-axis. }
    \label{fig:discr}
    \end{center}
\end{figure}

The so-obtained discretized region $\Omega$ is depicted in Figure~\ref{fig:discr}.

More specifically, we now have that:
\begin{eqnarray}
   X_j = (j-1) \; \Delta x  - x_{min} \nonumber \\
   Y_k = (k-1) \; \Delta y - y_{min}
\end{eqnarray}

for $j = 1, 2, \ldots, N_x$ and  $k = 1, 2, \ldots, N_y$.

Now, the possible functions in $\Omega$ can be expressed as the finite
set of vectors or discrete signals:
\begin{equation}
  \vec{f} = \left[  f_1 \; f_2 \;  \ldots \; f_{N_x}  \right]^T
\end{equation}

with $f_j$ taking values in the set $\left\{Y_k \right\}$ respectively
to the abscissae $X_j$.

It is assumed henceforth, typically with little loss of generality, that $x_{min}=-1$, 
$x_{max}=1$, $y_{min}=-1$,  $y_{max}=1$.

The total number of possible vectors in the discretized region $\Omega$
can now be calculated as being given as corresponding to the number
of permutations:
\begin{equation}
    N_T = N_x^{N_y}
\end{equation}

Henceforth, we identify each of the $N_T$ possible discrete signals 
(or functions) in $\Omega$ in terms of a respective label 
$n = 1, 2, \ldots, N_T$.  In case $N_y$ is relatively small, it is possible
to implement this association by deriving the discrete signal from its
respective label $n$ by first representing this value in radix $N_y$,
yielding the number $[p_{N_x-1} \; \ldots \; p_{1}  \; p_0]_{N_y}$, and
then making:
\begin{equation} \label{eq:assign}
   Y_{i+1} = p_{i} \, \Delta y + y_{min}
\end{equation}

for $i = 0, 1, \ldots, N_x-1$.

The difference between two discrete functions $\vec{f}$ and $\vec{h}$ 
can now be expressed in terms of the following root mean square error:
\begin{equation}  \label{eq:tau_d}
   \delta(\vec{f},\vec{h}) = \sqrt{ \frac{1}{x_{max}-x_{min}} \sum_{j = 1}^{N_x}  [f[X_j] - h[X_j] ]^2 }
\end{equation}

while the similarity between those functions can still be gauged by using
Equation~\ref{eq:simil}.   

In order to verify if a given function $\vec{f}$ can be approximated by a reference
function $g_i(x)$, we apply the linear least squares methodology (e.g.~\cite{CostaLeast}).
This approach provides the set of fit parameters (e.g.~the coefficients of a polynomial)
so as to minimize the error of the fitting as expressed by the sum of the square of the
differences between $\vec{f}$ and $g_i(x)$ (taken at the abscissae $X_f$).  For instance,
if $g_i(x)$ is a third degree polynomial and $N_x=5$, we first obtain the matrix:
\begin{equation}
   A = \left[  \begin{array}{c c c c}   
                1 & X_1 & X_1^2 & X_1^3 \\
                1 & X_2 & X_2^2 & X_2^3 \\
                1 & X_3 & X_3^2 & X_3^3 \\
                1 & X_4 & X_4^2 & X_4^3 \\
                1 & X_5 & X_5^2 & X_5^3 \\
           \end{array}  \right]    \nonumber
\end{equation}

and then express the respective coefficients in terms of the vector:
\begin{equation}
   \vec{p} = \left[  \begin{array}{c}   a_0 \; a_1 \; a_2 \; a_3    \end{array}  \right]^T \nonumber
\end{equation}

So that the fitting can be represented in terms of the following overdetermined system:
\begin{equation}
   \vec{f} =  A \; \vec{p}          
\end{equation}

The respective solution can be obtained in terms of the \emph{pseudo-inverse} of 
$A$ as:
\begin{equation}
   \vec{p} = (A^T A)^{-1} A^T \vec{f}               
\end{equation}

\begin{figure}[h!]  
\begin{center}
    \includegraphics[width=0.8\linewidth]{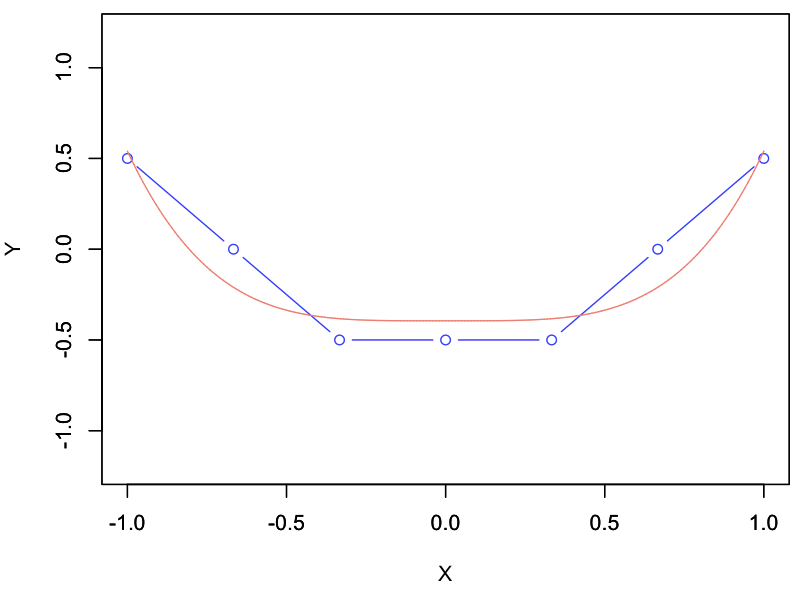}  \\  \vspace{0.1cm}
    \caption{Example of the linear least squares methodology for fitting a discrete
    signal $Y_i = f(X_i)$, with $N_x=7$ and $N_y = 5$, by a reference function of the type $a_1 x^4 + a_0$.
    The respectively obtained root mean square error was $\tau_d = 0.136$,
    implying a similarity of $\tau_s =  e^{-10 \, \tau_d} = 0.256$ (for $\alpha = 10$).}
    \label{fig:fitting}
    \end{center}
\end{figure}

\section{Discrete Signals Coverage} \label{sec:coverage}

The discretization of $\Omega$ implies that not all signals in $\vec{f}$ can be 
expressed  with full accuracy in terms of reference functions $\vec{g}_i$, so that it 
becomes important to adopt some difference tolerance $\tau_d$, or respectively 
associated similarity tolerance $\tau_s$.  Henceforth, every discrete signal $\vec{f}$ that
can be approximated by a reference function $g_i$ within a given tolerance $\tau$ will
be said to be \emph{adjustable} by that reference function.

It is important to keep in mind that, when a tolerance is allowed, more than one of the 
reference functions can be verified to provide a good enough (i.e.~with error smaller than 
the specified tolerance) approximation, in which case a same function $\vec{f}$ will
be identified as being adjustable by more than one reference function, which is
reasonable given that this actually happens in discrete domains.
However, in case the mapping is required to be made unique, it is possible
to keep only one of the fittings for each possible $\vec{f}$, such as that corresponding
to the smallest approximation error.   In this work, however, multiple adjustments will
be considered.

The sets $S_i(\tau_d)$, which are defined by $\tau_d$, will now contain a \emph{finite} number 
of discrete functions.  Thus, given a discrete signal $\vec{f}$ and a set of reference functions 
$g_i$, $i = 1, 2, \ldots, N$, the total number of adjustable signals $N_a$ can be expressed as:
\begin{equation}
  N_a = \sum_{k=1}^{N}S_k(\tau_d)
\end{equation}

We can now take the relative frequency of each reference 
function $g_i$ with respect to the whole of adjustable functions as:
\begin{equation}
  r(g_i, \tau_d) = \frac{\# \left\{ S_i(\tau_d) \right\}}{N_a}
\end{equation}

where $\# \left\{ S_i(\tau_d) \right\}$ corresponds to the \emph{cardinality} of the set 
$S_i(\tau_d)$.

This measurement, which is henceforth referred to as \emph{relative coverage},
can be used to compare the fitting potential of each of the considered
reference functions.

It is also possible to consider the following densities relative to the total number of functions in 
$\Omega$ as:
\begin{equation}  \label{eq:q}
  q(g_i, \tau_d) = \frac{\# \left\{ S_i(\tau_d) \right\}}{N_T}
\end{equation}

In case only one fitting is associated to each possible discrete signal $\vec{f}$ in $\Omega$, 
we will have that $0 \leq q(g_i, \tau_d) \leq 1$ and that $\sum_{k=1}^N q(g_i, \tau_d) = 1$.  
This can be achieved by considering the sets $\tilde{S_i} = S_i - \bigcup_{k=1}^{N} S_{k \neq i}$
instead of $S_i$ in Equation~\ref{eq:q}.
Otherwise, this measurement may take values larger than 1 and we will also have that
$\sum_{k=1}^N q(g_i, \tau_d) \geq 1$, indicating that the possible discrete 
functions in $\Omega$ is being covered in excess.

The relative density $q(g_i, \tau_d)$, henceforth called the \emph{coverage index} of $g_i$
provides a means to quantify of how well the reference function $g_i$ \emph{covers} the 
discrete signals in the given region $\Omega$ and resolution $\tau_d$.   Larger values of 
$q(g_i, \tau_d)$ will typically be observed when $\tau_d$ is increased (or $\tau_s$ is decreased).   

Also, observe that the above relative densities also depend on the choice of the discretization
resolutions $\Delta x$ and $\Delta y$, with $\# \left\{ S_i(\tau_d) \right\}$ increasing
substantially with $N_x$ and $N_y$.

\section{Discrete Functions Adjacency}

While the relative densities $r(g_i,\tau_d)$ can provide interesting insights about
the generality of each considered reference function $g_i$, these measurements
can provide no information about the proximity or interrelationship between the discrete 
functions $\vec{f}$ as fitted by a set of reference functions $g_i$, $i = 1, 2, \ldots, N$.
However, it is possible to quantify the proximity between all the possible discrete
functions in $\Omega$ in terms of some distance between the respective vectors and then
define links between the pairs of functions that have respective distances smaller than
a given threshold $L$.

Consider the Euclidean distance between two discretized 
functions $\vec{f^{[i]}}$ and $\vec{f^{[j]}}$ in the region $\Omega$ as:
\begin{equation}
   \omega \left( \vec{f^{[i]}},\vec{f^{[j]}} \right) = \sqrt{\sum_{k=1}^{N_x}  \left[ f^{[i]}_k - f^{[j]}_k \right]^2 }
\end{equation}

The whole set of Euclidean distances between every possible pair of functions in 
a given $\Omega$ can then be represented in terms of the following distance matrix:
\begin{equation}
   W_{i,j} =  w \left( \vec{f^{[i]}},\vec{f^{[j]}} \right)
\end{equation}

The symmetric matrix $W$ can be immediately understood as providing the strength of the
links between the nodes of a graph, each of these nodes being associated to one of the
possible $N_T$ discrete functions in a given $\Omega$.  However, such a graph would
express the \emph{distances} between functions, not their \emph{proximity}.  Though 
these distances could be transformed into similarity measurements by adopting an 
expression analogous to Equation~\ref{eq:simil}, therefore yielding a weighted 
respective graph, in this work we adopt the alternative approach of understanding two
discrete functions as being \emph{adjacent} provided the respective Euclidian distance
as defined above is smaller or equal to a given threshold $L$.

Overall, obtaining the transition network for a set of reference
functions $g_i(x)$ and a respective discrete region $\Omega$, with $N_x$ and $N_y$,
involves the following 3 main processing stages:

\begin{itemize}
  \item Assign a label $n$ to each of the $N_T = N_x^{N_y}$ possible discrete signals in $\Omega$;
  \item For each value $n = 1, 2, \ldots, N_T$, obtain the respective function 
           $\vec{f}_n = [Y_{N_x-1}, \ldots, Y_2, Y_1]$
           by using Equation~\ref{eq:assign} and apply least square approximation respectively
           to each of the reference functions $g_i(x)$, $i=1, 2, \ldots,N$.  In case the similarity between
           $\vec{f}_n$ and $g_i$ as obtained by applying Equations~\ref{eq:tau_d} and then ~\ref{eq:simil}, 
           is larger or equal to $\tau_s$, assign a respective node with label $n$, 
           also incorporating the type $i$ of the respective approximating function $g_i(x)$;
  \item Interconnect all pairs of nodes obtained in the previous step which have Euclidean distance
           smaller than $L$, therefore yielding the transition network  $\Gamma$.
\end{itemize}

It is also important to keep in mind that one so obtained transition network can be understood as
constraining the overall adjacency network between all possible functions $\vec{f}$ in $\Omega$
so that only the nodes associated to cases that can be adjusted with good accuracy by a
respective reference function $g_i$ are maintained.  In brief, the transition network therefore
provides a representation of the adjacency between the possible discrete functions that can 
be adjusted by the reference functions.

\section{Optimized Transitions and Random Walks}

The derivation of the transition network $\Gamma$ respective to a set of reference functions and a discrete
region $\Omega$ paves the way to several interesting analysis and simulations, some of which are
discussed in this section.

One first interesting possibility is, given two functions $\vec{f}_i$ and $\vec{f}_j$ in $\Gamma$,
to identify the \emph{shortest paths} between the respective nodes.  We mean paths in the 
plural because it may happen that more than one shortest path exist between any two nodes of 
a network.  Each of these obtained shortest paths indicate the smallest number of successive 
transitions from $\vec{f}_i$ to $\vec{f}_j$  that are necessary to take one of those functions 
into the other (or vice-versa) while using only instances of the considered reference functions.  
This result is potentially interesting for several applications, including implementing optimal 
controlling dynamics underlain by the reference functions, or optimal morphing between two 
or more signals underlain by the respectively considered reference functions.  

Given a transition network $\Gamma$ and all the shortest paths between its pairs of nodes, it also
becomes interesting to consider statistics of the length of those paths, such as their average and
standard deviation, which can provide interesting information about the overall potential of the 
reference functions for implementing optimal transitions and morphings as mentioned above.

Another interesting approach considering a transition network consists in performing 
\emph{random walks} (e.g.~\cite{Justin}) along its nodes.   Several types of random walks can 
be adopted, including uniformly random and preferential choice of nodes according to several local topological properties of the network nodes, such as degree and clustering coefficient.  
These random walks can be understood as implementing respective
types of dynamics in the network.  For instance, a random walk with uniform transition probabilities 
is intrinsically associated to diffusion in the network.  In this manner, random walks on transition
networks provide means for simulating and characterizing properties related to dynamics involving
transition between the discrete signals in $\Gamma$.

Yet another interesting perspective allowed by the derivation of the transition matrices $\Gamma$ concerns studies involving betweenness centrality (e.g.~\cite{surv_meas})
or accessibility (e.g.~\cite{accessibility}) of edges and nodes in $\Gamma$, which can
complement the two aforementioned analyses.  For instance, it could be interesting to use the 
accessibility to identify the discrete signals in $\Omega$, as underlain by the reference 
functions $g_i(x)$, leading to the largest and smallest number of nodes, therefore providing 
information about the role of those nodes regarding influencing or being influenced by other nodes.  
The accessibility measurement can also be applied in order to identify the center and periphery of 
the obtained transition networks~\cite{borders_access}.

\section{Case Example 1: Power Functions}

\begin{figure}[h!]  
\begin{center}
    \includegraphics[width=0.9\linewidth]{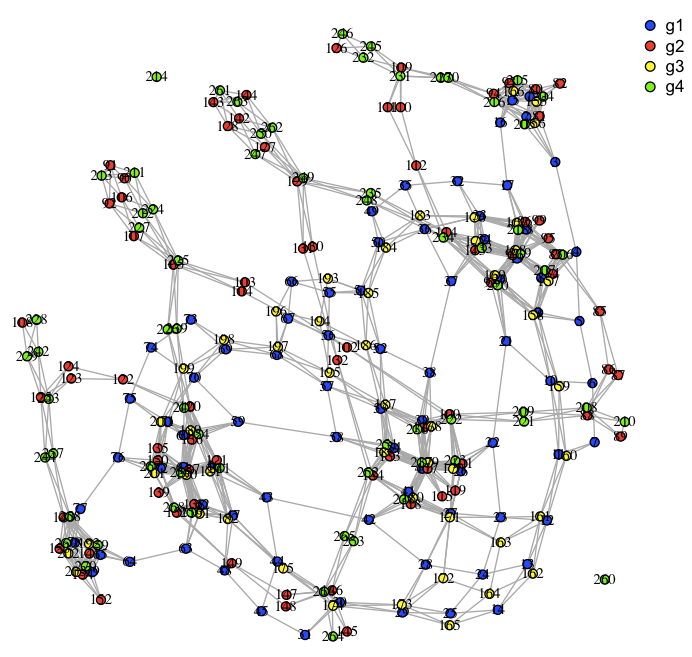}  \\  \vspace{0.1cm}
    \caption{Visualization, by using the Fruchterman-Reingold method, of the transition 
    network obtained for the reference power functions in Eq.~\ref{eq:four_ref},  $N_x=5$ and
    $N_y=5$, assuming $\tau_s = 0.2$ and $L = 0.6$. The colors indicate,
    according to the legend, the respective type of power function approximating the 
    discrete signals.  The two semiplanes of the bilateral symmetry corresponds to
    the sign of the coefficients $a^i_1$.  The five main clusters of nodes correspond to the 
    constant functions $a^i_0 = -1, -0.5, 0, 0.5, 1$.  Observe the hubs at the center of each
    of the 5 clusters of nodes.  See text for more information. }
    \label{fig:transition_5}
    \end{center}
\end{figure}

This section presents a case example of the proposed methodology assuming the four
power functions in Equation~\ref{eq:four_ref}.   First, we consider the region $\Omega$  as
being sampled by $N_x = 5$ abscissae values and $N_y = 7$ coordinate samples,
assuming $\tau_s = 0.2$, $\alpha = 10$, and $L=0.6$.  The resulting transition network is depicted
in Figure~\ref{fig:transition_5}, as visualized by the Fruchterman-Reingold 
methodology~\cite{Fruchterman}.

Several remarkable features can be identified in the obtained transition network.
First, we find the nodes organized according to a well-defined bilateral symmetry, which can
be verified to correspond to the sign of the coefficients $a^i_1$, $i=1, 2, 3, 4$.  In addition,
the nodes corresponding to approximations by the power functions $g_1$ and $g_3$, 
both of which presenting odd parity, tend to be adjacent one another, with a similar
tendency being observed for the nodes respective to the evenly symmetric power functions 
$g_2$ and $g_4$.   Five main clusters of nodes can also be identified along the diagonal of
the figure running from bottom-left to top-right, each of which with a respective central hub. 
These hubs correspond to the constant functions  $a^i_0 = -1, -0.5, 0, 0.5, 1$ which, as
discussed in Section~\ref{sec:adjacency}, represent transition points of the adopted set of 
reference functions and $\Omega$.  As could be expected, these hubs and surrounding
clusters of nodes, are characterized by the presence of all the four types of considered
power functions.  The other, smaller, clusters of nodes are associated to transition
points allowed by the  adoption of a non-null tolerance, and possibly reflect the intrinsic
structure of the discrete space $\Omega$.

\begin{figure}[h!]  
\begin{center}
    \includegraphics[width=0.9\linewidth]{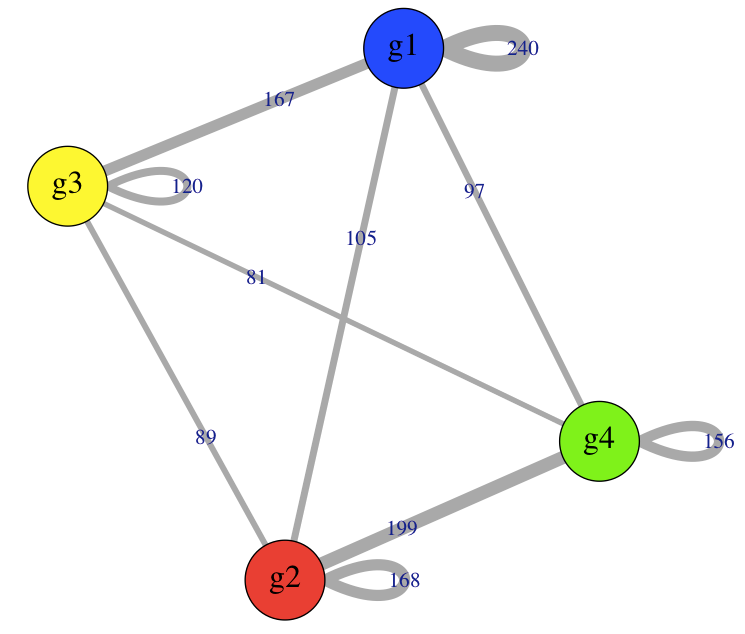}  \\  \vspace{0.1cm}
    \caption{The reduced version of the transition network in Fig.~\ref{fig:transition_5}.
    Each of the 4 nodes correspond to a respectively indicated type of power
    function, while the links between each pair of nodes accumulate all 
    connections between the original subsumed nodes.}
    \label{fig:reduced}
    \end{center}
\end{figure}

It is also possible to derive a \emph{reduced} version of the above transition network.
Basically, all nodes associated to each of the 4 categories of nodes (i.e.~the adopted
4 reference functions) are subsumed by a respective node, while the interconnections
between all the original nodes are also collected into the links between the agglomerated
nodes.   Figure~\ref{fig:reduced} illustrates the reduced version of the transition network
in Figure~\ref{fig:transition_5}.

The obtained reduced graph corroborates the predominance of transitions between
odd ($g_1$ and $g_3$)  and even  ($g_2$ and $g_4$)  power functions.
In addition, we also have that the largest number of transitions is observed between
instances of $g_1$, and that the smallest number of transitions takes place between
instances of $g_3$ and $g_4$.  The largest number of transitions between odd and
even functions take place between $g_1$ and $g_2$.

\begin{figure*}[h!]  
\begin{center}
    \includegraphics[width=0.3\linewidth]{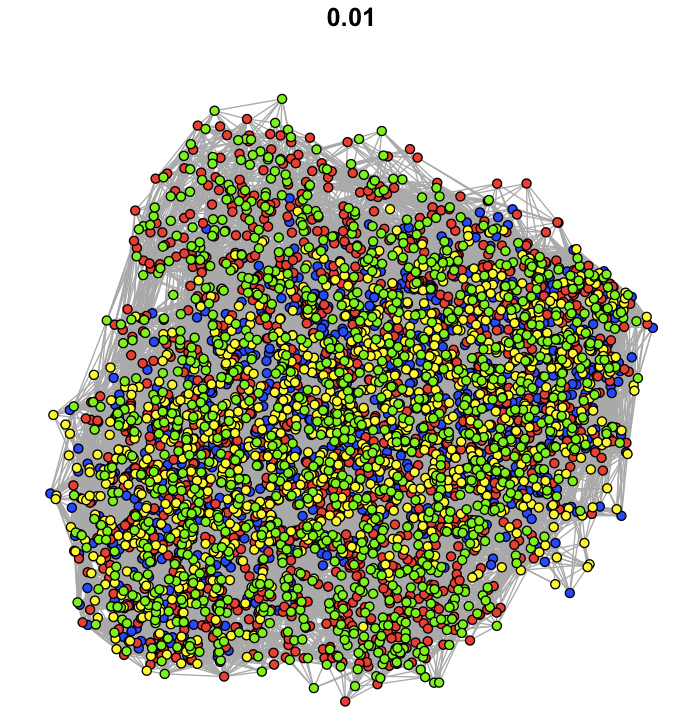}  
    \includegraphics[width=0.3\linewidth]{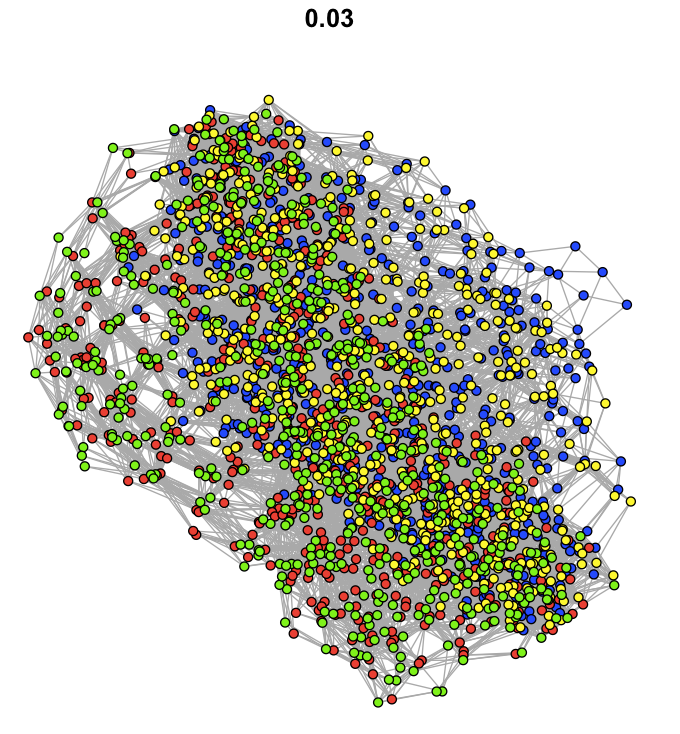}  
    \includegraphics[width=0.3\linewidth]{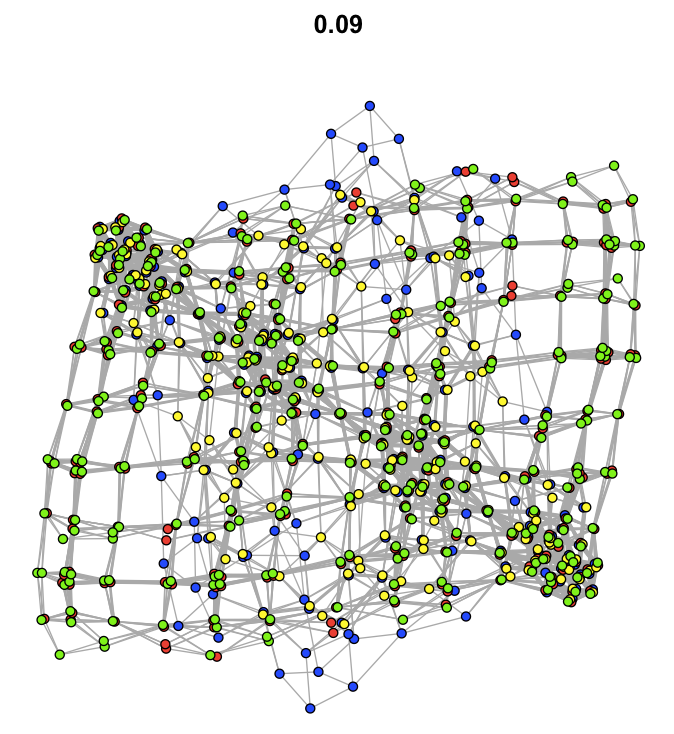}  \\   \vspace{0.7cm}
    \includegraphics[width=0.3\linewidth]{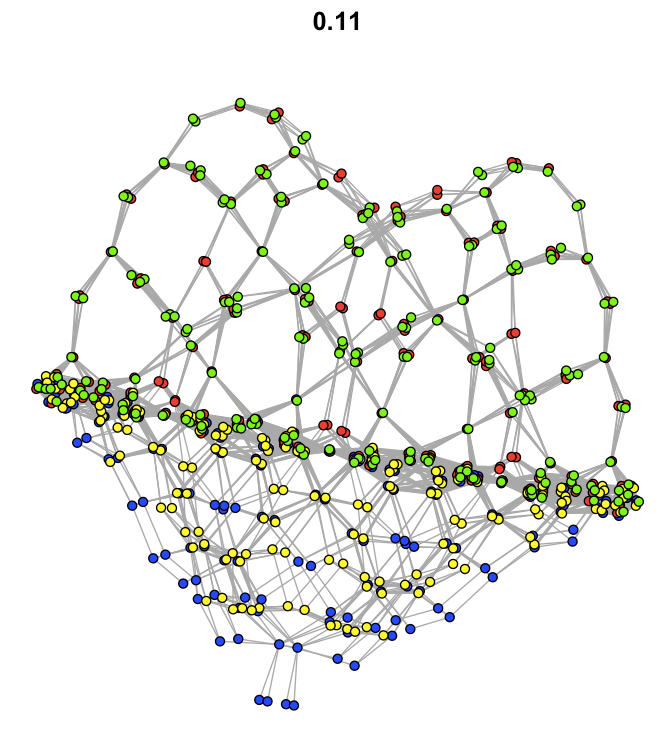}  
    \includegraphics[width=0.3\linewidth]{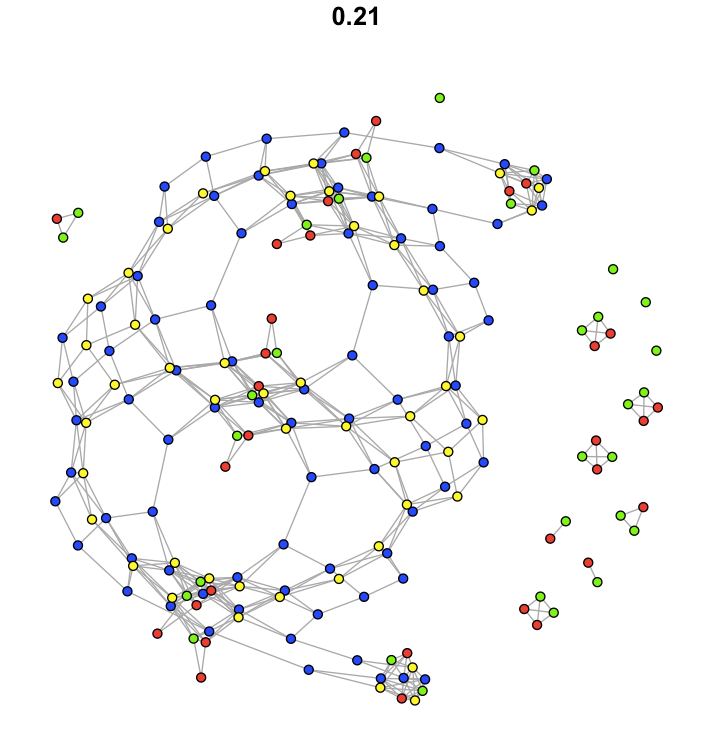}  
    \includegraphics[width=0.3\linewidth]{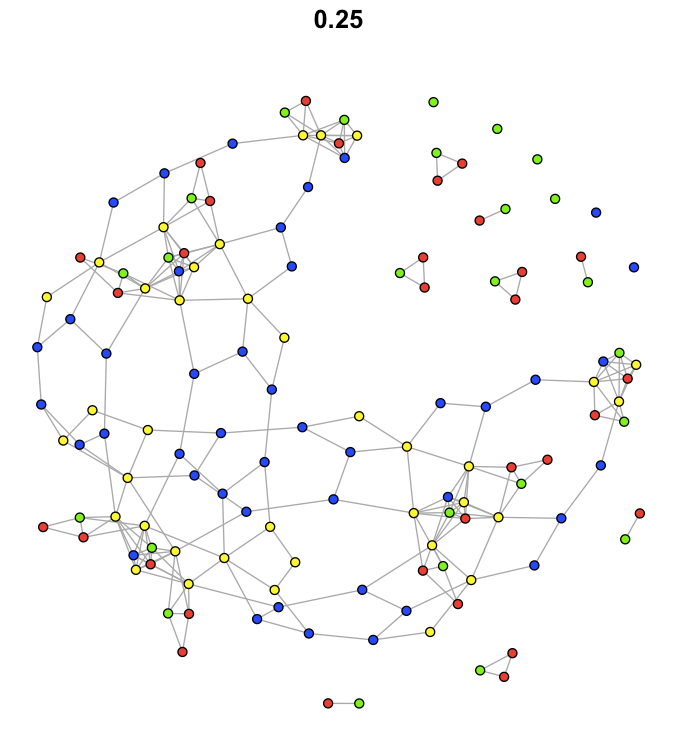}  \\  \vspace{0.7cm}
    \includegraphics[width=0.3\linewidth]{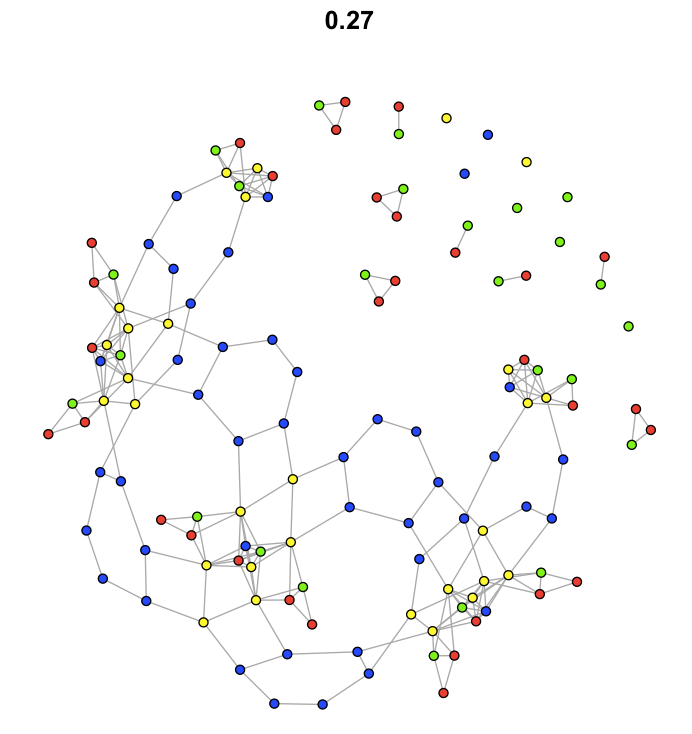}  
    \includegraphics[width=0.3\linewidth]{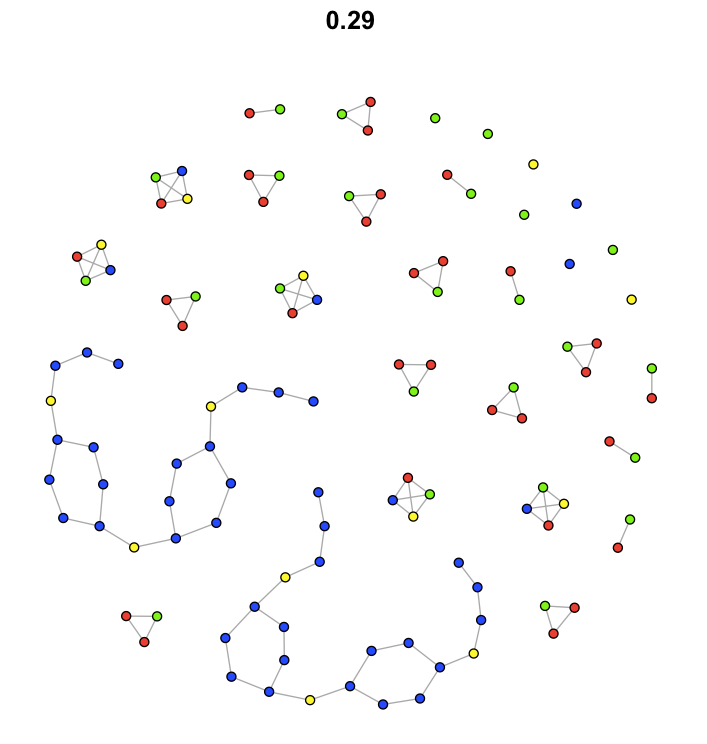}  
    \includegraphics[width=0.3\linewidth]{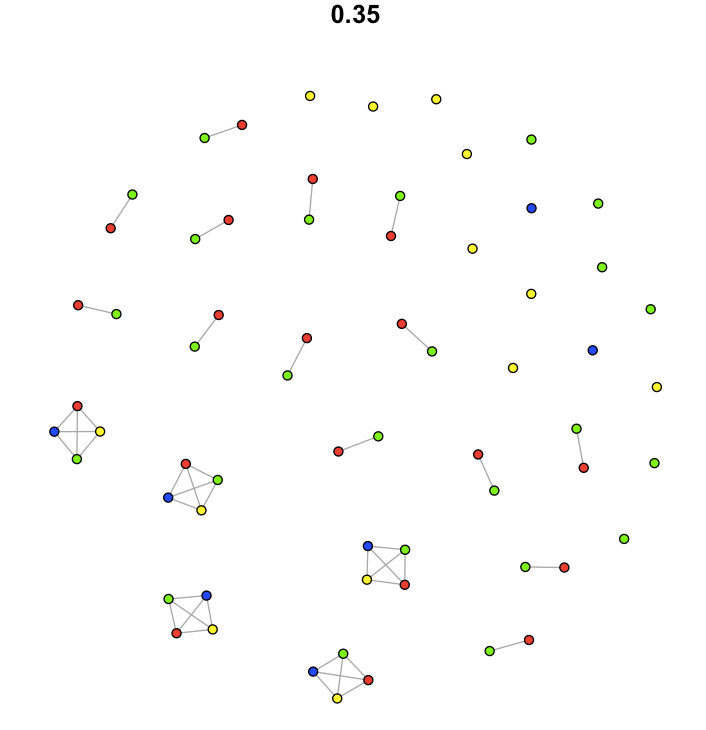}  
    \caption{Additional examples of transition networks obtained for the same configuration
    used in the previous example, but with respect to several other values of $\tau_s$, as
    respectively indicated in each network.  The colors follow the same convention
    as in Fig.~\ref{fig:transition_5}.  Network visualizations obtained by using 
    the Fruchterman-Reingold method.}
    \label{fig:several}
    \end{center}
\end{figure*}

One interesting question regards to what an extent the transition networks may vary with
respect to distinct values of the tolerance $\tau_d$ or $\tau_s$.  Figure~\ref{fig:several}
depicts 9 additional transition networks obtained for the same configuration adopted in the
previous example, with respect to several different values of $\tau_s$ respectively indicated
above each network.

As illustrated in Figure~\ref{fig:several}, the size and connectivity of the transition network
decreases steadily with $\tau_s$, and several markedly distinct types of networks, most of which
presenting bilateral symmetry, are respectively observed.  Given that the more generalized
ability of the power functions to adjust the discrete signals when larger tolerance values are
allow (i.e.~small values of $\tau_s$), the initial networks tend to present a more widespread 
and uniform interconnectivity.   Observe also that the networks
split into two or more connected components for values of $\tau_s$ larger than
approximately $0.2$.

\begin{figure}[h!]  
\begin{center}
    \includegraphics[width=0.9\linewidth]{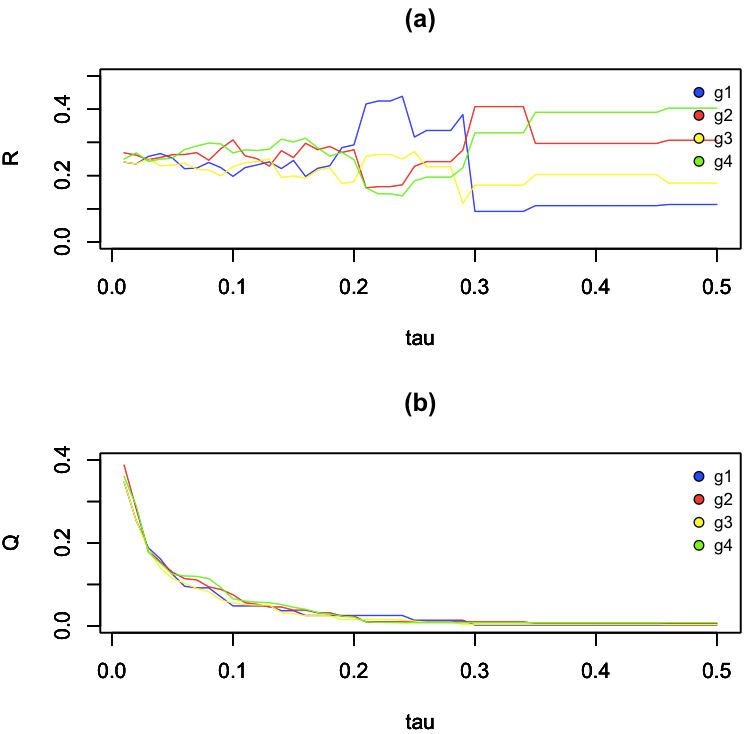}  \\  \vspace{0.1cm}
    \caption{The relative coverage (a) and coverage index (b) for the case example 1
    with $N_x=5$ and considering the reference functions in Equation~\ref{eq:four_ref},
    as obtained for similarity tolerance values $\tau_s = 0.01, 0.02, \ldots, 0.5$.}
    \label{fig:coverage_5}
    \end{center}
\end{figure}

The relative coverage and coverage index (see Section~\ref{sec:coverage}) of the four considered
power functions for $\tau_s = 0.01, 0.02, \ldots, 0.5$ are shown in Figures~\ref{fig:coverage_5}(a)
and (b), respectively.

Similar values of relative coverage can be observed for the four power functions, with 
oscillations along $\tau_s$ that tend to increase from left to right in Figure~\ref{fig:coverage_5}(a),
up to a point, near $\tau_s =0.35$, where the relative coverages become nearly constant
and markedly distinct between the 4 considered types of reference functions.

As expected, the coverage index decreased steadily with $\tau_s$ for all the four considered
reference functions, also presenting values similar. Observe that only a small percentage of
the possible discrete signals are adjustable at $\tau_s = 0.2$.

\begin{figure}[h!]  
\begin{center}
    \includegraphics[width=1\linewidth]{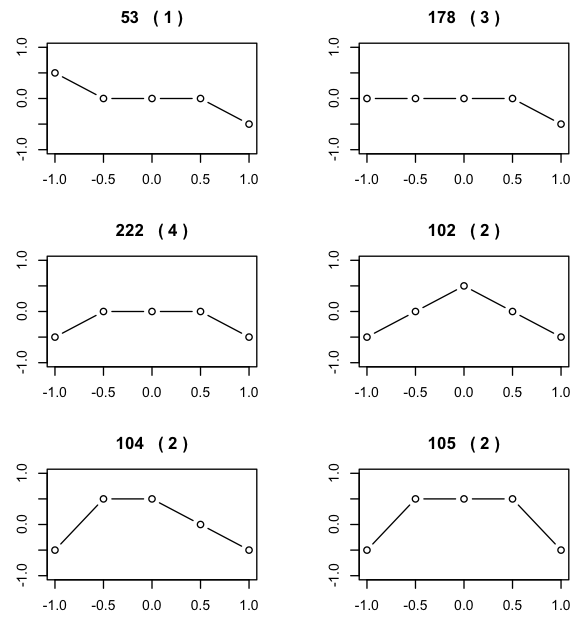}  \\  \vspace{0.1cm}
    \caption{The minimal sequence of transitions in the transition network in Figure~\ref{fig:transition_5},
    assuming $N_x=5$ and $N_y = 5$,
     leading from signal $n=53$ to signal $n=105$.  The numbers within parenthesis indicate the
     type of respectively fitted power function ($1 = g_1$, $2 = g_2$, $3 = g_3$, and $4 = g_4$).}
    \label{fig:shortest_5}
    \end{center}
\end{figure}

Let's now consider the shortest path between two functions in the above transition network.  
Figure~\ref{fig:shortest_5} illustrates the shortest sequence of transitions between the functions respectively identified by the numbers $53$ and $105$.

\begin{figure*}[h]  
\begin{center}
    \includegraphics[width=1\linewidth]{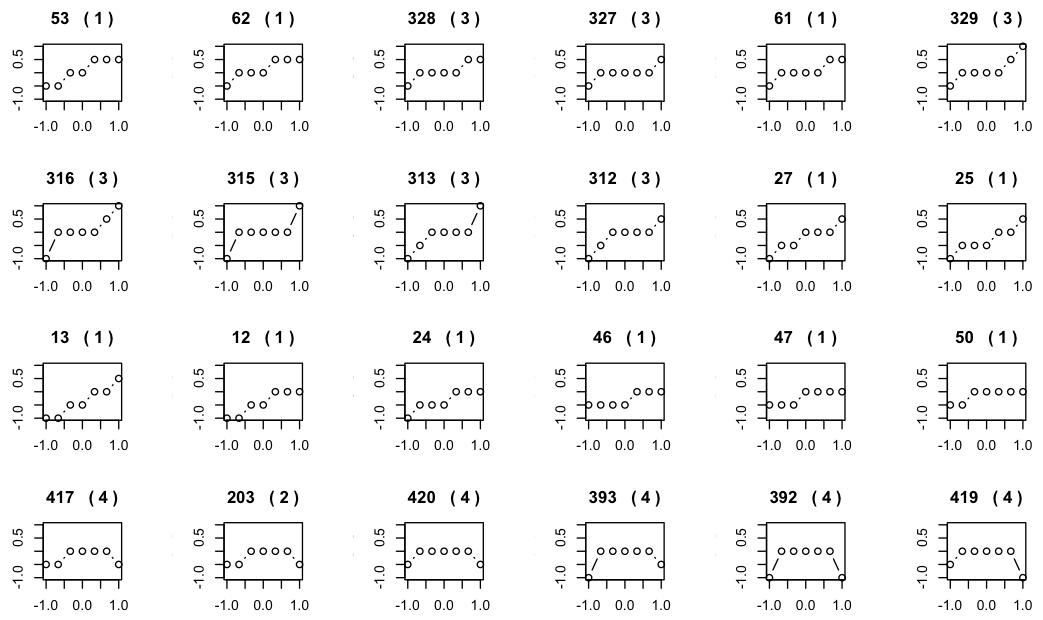}  \\  \vspace{0.1cm}
    \caption{One of the many possible random walks with 23 steps in the transition network
    shown in Figure~\ref{fig:transition_5}, considering self-avoiding uniform transition probabilities.
    Observe the incremental change implemented in the involved discrete signals at each
    successive step.  The numbers within parenthesis indicate the
     type of respectively fitted power function ($1 = g_1$, $2 = g_2$, $3 = g_3$, and $4 = g_4$).}
    \label{fig:rand_5}
    \end{center}
\end{figure*}

Figure~\ref{fig:rand_5} shows the first  23 steps of a possible self avoiding random walk along the 
above transition network, starting at signal $n = 53$.

Self avoiding operation was adopted in not to repeat nodes.  Observe the relatively
smooth transition, involving minimal modifications of the discrete signals, along each of the
implemented transitions.  

\begin{figure}[h!]  
\begin{center}
    \includegraphics[width=0.9\linewidth]{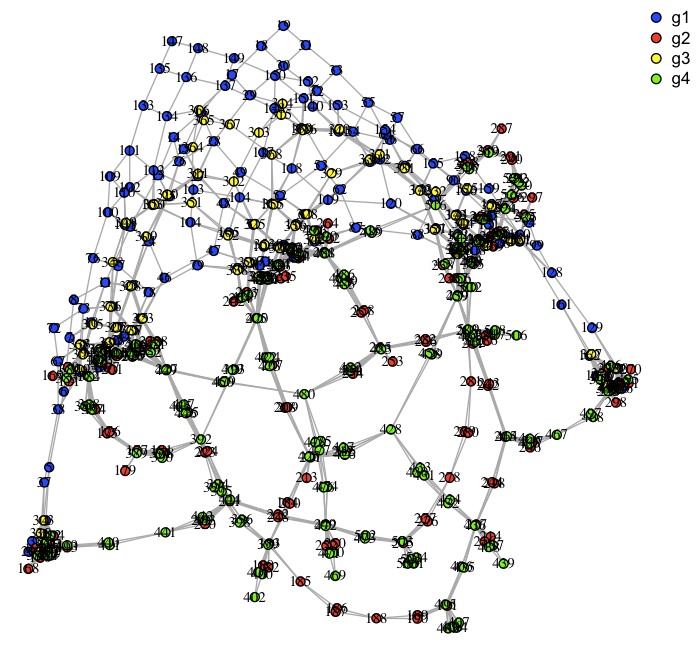}  \\  \vspace{0.1cm}
    \caption{Visualization, by using the Fruchterman-Reingold method, of the transition 
    network obtained for the reference power functions in Eq.~\ref{eq:four_ref},  $N_x=7$ and
    $N_y=5$, assuming $\tau_s = 0.2$ and $L = 0.6$. The colors indicate,
    according to the legend, the respective type of power function approximating the 
    discrete signals.   See text for more information.}
    \label{fig:transition_7}
    \end{center}
\end{figure}

Figure~\ref{fig:transition_7} depicts the transition network obtained for the same situation above,
but now with $N_x=7$ instead of $N_x=5$.  

The resulting transition network again presents several interesting features.  As before, we have
the bilateral symmetry corresponding to the sign of the coefficients associated to the $x$-term.
In addition, clusters and respective central hubs have again be obtained, corresponding to the
constant (null) transition functions as observed before.  However, unlike the network obtained
for $N_x=5$, now we most of the nodes separated also along the up-down orientation,
corresponding to interactions between blue-yellow (up) and red-green (down).  These two
portions of the transition network can therefore be understood as being directly associated
to the odd/even parity of the involved reference functions.   Of particular interest is the fact
that the discrete signals associated to the blue nodes, associated to the reference function  
$g_1(x) = a^1_1 x + a^1_0$, define a relatively regular pattern of interconnection that is
markedly distinct to the more sequential pattern of interconnections observed for the 3
other reference functions.  Observe that this transition network also incorporates several
\emph{handles}, corresponding to relatively long sequences of links~\cite{tails_handles}.
Such sequences are associated to incrementally distinct instances of the same type of reference
function, as discussed in Section~\ref{sec:adjacency}.

\section{Case Example 2: Polynomials}

While the previous case example assumed power functions containing only two terms,
we now address the more general situation where only one complete polynomial of order 
$P$ is adopted as reference function, i.e.:
\begin{equation}
   g_1 = a_P x^P + \ldots + a_2 x^2 + a_1 x + a_0
\end{equation}

\begin{figure}[h!]  
\begin{center}
    \includegraphics[width=0.9\linewidth]{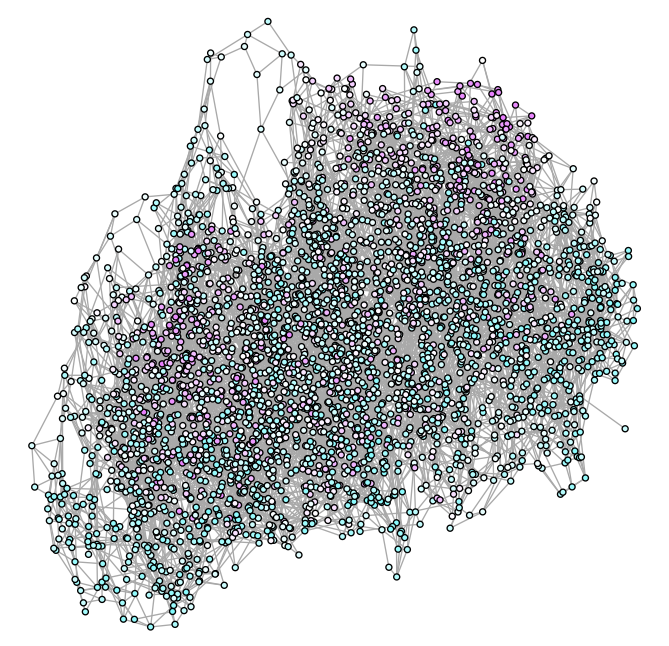}  \\  \vspace{0.1cm}
    \caption{Visualization, by using the Fruchterman-Reingold method, of
    the transition network obtained for a complete polynomial of order $P=4$
    as single reference function, and $N_x = 7$, $N_y = 5$, $\tau_s = 0.4$, $\alpha = 10$,
    and $L = 0.6$.  The colors are assigned so that increasing values 
    are represented from cyan to magenta color tones.}
    \label{fig:poly}
    \end{center}
\end{figure}

Figure~\ref{fig:poly} illustrates the transition network obtained for the above polynomial
reference function assuming $P = 4$, $N_x = 7$, $N_y = 5$, $\tau_s = 0.4$, $\alpha = 10$,
and $L = 0.6$.

Interestingly, a completely different topology is now observed for the polynomial transition network
as compared to the previous the networks respective to power functions.  The main distinguishing
features are two: (i) a much larger number of nodes are now observed; and (ii)
their interconnectivity is much more uniform, without present of well-defined heterogeneities
such as clusters, hubs, tails or handles.     All these properties can be understood as being
consequence of the substantially higher flexibility that a complete polynomial function has 
for adjusting signals as compared to those of the more specific power functions considered 
previously in this work.  As a consequence of this enhanced adjusting property, many more 
discrete signals could be fitted with reasonably accuracy, hence the larger network size 
obtained.  The observed uniformity of connections also follows from the flexibility of complete
polynomials, as they cater for many more transition points
corresponding to the larger number $P$ of involved terms and parameters.

\section{Case Example 3: Hybrid Functions}

As with power functions and polynomials, also sinusoidal functions are extensively applied
in mathematics, physics, and science in general, constituting the basic components of
the flexible Fourier series.  The third case example considered in the
present work adopts a set of reference functions containing two power functions and two
sinusoidals, more specifically:
\begin{eqnarray}
    g_1(x) = a^1_1 x + a^1_0 \nonumber \\
    g_2(x) = a^2_1 x^2 + a^2_0 \nonumber \\
    g_3(x) = a^3_1 \, sin(3 x) + a^3_0 \nonumber \\
    g_4(x) = a^4_1 \, sin(5 x)  + a^4_0   \label{eq:hybrid}
\end{eqnarray}

\begin{figure}[h!]  
\begin{center}
    \includegraphics[width=1\linewidth]{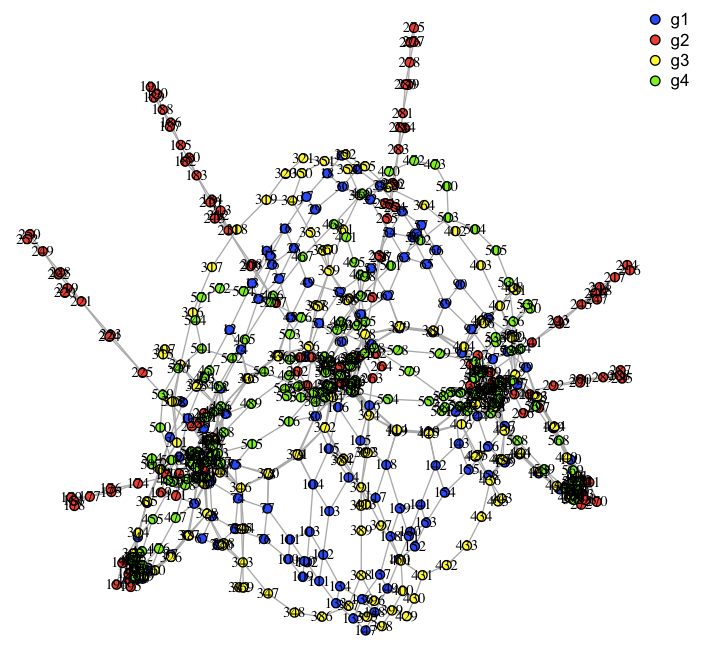}  \\  \vspace{0.1cm}
    \caption{Visualization, by using the Fruchterman-Reingold method, of the transition 
    network obtained for the reference hybrid (two power and two sinusoidal)
    functions in Eq.~\ref{eq:hybrid}, for  $N_x=7$ and
    $N_y=5$, $\tau_s = 0.2$ and $L = 0.6$. The colors indicate,
    according to the legend, the respective type of power function approximating the 
    adjustable discrete signals.  In addition to the 5 clusters observed in the previous
    examples involving power functions, now also tails and handles are obtained.
    See text for additional discussion.}
    \label{fig:hybrid}
    \end{center}
\end{figure}

Figure~\ref{fig:hybrid} illustrates the transition network obtained for the above hybrid reference
functions assuming $P = 4$, $N_x = 5$, $N_y = 7$, $\tau_s = 0.2$, $\alpha = 10$, and $L = 0.6$.

A particularly interesting structure is observed for this example.  First, the five
main clusters, corresponding to respective constant transition points, are
again observed in analogous manner with the other examples involving power
functions.  bilateral symmetry is again observed, being related to the sign of the
coefficients $a^i_1$, $i = 1, 2, 3, 4$.   Now, surrounding those clusters of nodes, which incorporate 
all four types of reference functions, we also discern a
relatively regular subnetwork involving the first order power $g_1$ (blue) and the
lower frequency sinusoidal $g_3$ (yellow), both of which have odd parity.  These
nodes also tend to form handles at the border of the obtained transition network.

The nodes associated to the second order power function $g_2$ (red) results mostly distributed
along the six projecting tails at the periphery of the network, which correspond to
incremental instantiations of the same type of function.  Contrariwise, the
nodes corresponding to discrete signals adjustable by the high frequency 
sinusoidal $g_4$ (green) are found concentrated in the three most central clusters of
nodes of the network, despite the fact that both $g_2$ and $g_4$ share even
parity.

In order to study the effect of extending a set of reference functions on the topology of the
respectively defined transition network, we incorporate two additional power functions, respective to 
third and forth orders, into the set of reference functions adopted in the previous example
(Eq.~\ref{eq:hybrid}), yielding the following extended set of reference functions:
\begin{eqnarray}
    g_1(x) = a^1_1 x + a^1_0 \nonumber \\
    g_2(x) = a^2_1 x^2 + a^2_0 \nonumber \\
    g_3(x) = a^3_1 x + a^3_0 \nonumber \\
    g_4(x) = a^4_1 x^2 + a^4_0 \nonumber \\
    g_5(x) = a^5_1 \, sin(3 x) + a^5_0 \nonumber \\
    g_6(x) = a^6_1 \, sin(5 x)  + a^6_0   \label{eq:hybrid_b}
\end{eqnarray}

\begin{figure}[h!]  
\begin{center}
    \includegraphics[width=1\linewidth]{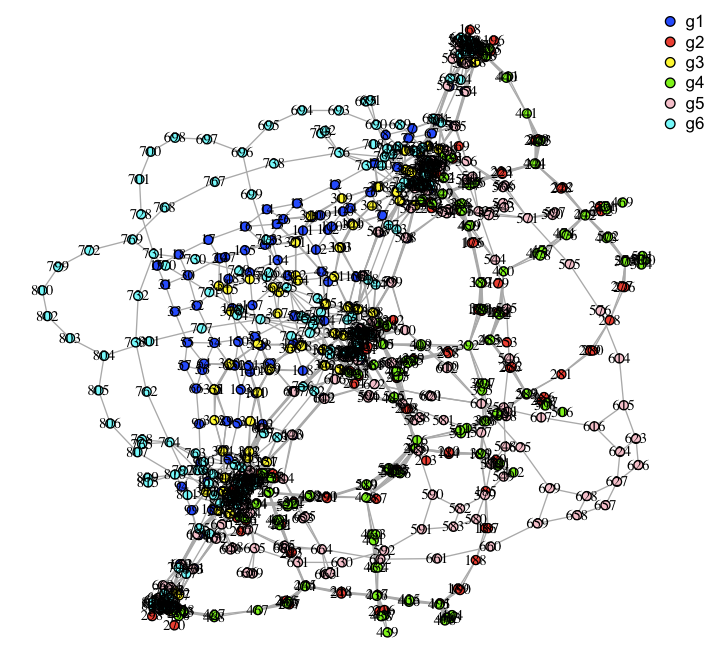}  \\  \vspace{0.1cm}
    \caption{Visualization, by using the Fruchterman-Reingold method, of the transition 
    network obtained for the second case of reference hybrid (four power and two sinusoidal)
    functions as in Eq.~\ref{eq:hybrid_b}, for  $N_x=7$ and
    $N_y=5$, $\tau_s = 0.2$ and $L = 0.6$. The colors indicate,
    according to the legend, the respective type of power function approximating the 
    adjustable discrete signals.  In addition to the 5 clusters observed in the previous
    examples involving power functions, now also tails and handles are obtained.
    See text for additional information.}
    \label{fig:hybrid_b}
    \end{center}
\end{figure}

Figure~\ref{fig:hybrid_b} illustrates the transition network obtained for the above hybrid reference
functions assuming $P = 4$, $N_x = 5$, $N_y = 7$, $\tau_s = 0.2$, $\alpha = 10$, and $L = 0.6$.

It is particularly interesting to contrast the obtained transition network in Figure~\ref{fig:hybrid_b}
with the networks in Figure~\ref{fig:transition_7}, obtained for four power functions, and that in
Figure~\ref{fig:hybrid}, which considers two power functions and two sinusoidal functions.
Therefore, it could be expected that network in Figure~\ref{fig:hybrid_b}, respective to the union 
of the reference functions in the two aforementioned sets, inherits some of their respective 
topological features.

Indeed, the network in Figure~\ref{fig:hybrid_b} incorporates some features from both the 
related structures.   First, we again observe the bilateral symmetry also common to those 
previous networks.  In addition, the obtained network can be understood, to a good extent, 
to the structure in Figure~\ref{fig:transition_7}
to which peripheral subnetworks corresponding to the two sinusoidals ($g_5$ in pink and $g_6$ in cyan)
have been incorporated, being characterized by several respective handles.  Also of particular interest
is the fact of the tails in Figure~\ref{fig:hybrid_b} being assimilated into the inner structure of the
network.

\section{Concluding Remarks}

Functions can be understood as essential mathematical concepts, being widely used both from the
theoretical and applied points of view in science and technology.  As a consequence of their great
importance, whole areas of mathematics and other major areas have been dedicated to their study and 
applications, including calculus, mathematical physics, linear algebra, functional analysis, 
numerical methods, numerical analysis, dynamic systems, and signal processing, to name but a few 
examples.

The present work situates at the interface between several of these areas, also encompassing other
areas, including network science, computer graphics, and shape analysis.  More specifically, we aimed
at developing the issue of how well all possible signals in a given region $\Omega$ correspond to
instances of a given set of reference functions.  Given that infinite sets of adjustable functions would be
obtained when working with continuous functions, we focused instead on addressing the aforementioned problem in discrete regions, leading to finite sets of adjustable functions to be obtained.  
In particular, if the region $\Omega$ is sampled by $N_x \times N_y$ values, the total number
of possible discrete signals in that region is necessarily equal to $N_x^{N_y}$.  The adoption of 
discrete signals also paves the way to verify if each of them can be adjusted, given a 
pre-specified tolerance, as instances of the reference parametric functions by using the least 
linear squares methodology.

Having identified the sets of adjustable discrete signals respectively to each of the adopted reference
functions, it becomes possible not only to study their relative density, but to approach the particularly
interesting issue of transitions between adjacent functions, yielding respective transition networks.
The adjacency between two functions, as understood in this work, was first characterized with respect
to continuous parametric functions as corresponding to respective instances leading to the identity
between the two functions, being subsequently adapted to discrete signals and functions by taking into 
account the Euclidian distances smaller than a specified threshold $L$.

A number of interesting possible investigations can then be performed with basis on these obtained
networks, including studies of optimal sequence of transitions, random walks potentially associated
to dynamical systems, as well the identification of particularly central signals in terms of betweenness
centrality and accessibility.

The potential of the reported concepts and methods were then illustrated with respect to three 
case examples respective to: (i) four power functions; (ii) a single complete polynomial of
forth order; and (iii) two sets of  hybrid reference functions involving combinations of  power 
functions and sinusoidals.  

As expected, the coverage index decreased steadily as $\tau_s$ increased, while the four
power functions presented similar potential for adjusting the discrete functions in the assumed
region $\Omega$.  

In addition, the obtained transition networks gave rise to a surprising diversity of topologies,
including combinations o modularity and regularity, as well as hubs, handles and tails.
Several of the networks also were characterized by symmetries which have been found
to be related to the sign of the reference function coefficients, as well as their parity.
The power functions and sinusoidals were found to lead to quite distinct patterns of
interconnectivity in the resulting transition networks, wth the latter leading to peripheral
handles.

The intricate and diverse patterns of topological structure obtained for the transition
networks are also influenced by the discrete aspects of the lattice underlying $\Omega$.
For instance, most of the case examples involving power and sinusoidals for $N_x=5$
were found to incorporate five clusters of nodes associated to the null
discrete transition.  Other topological heterogeneities of the obtained networks
are also related to specific anisotropies of the lattice, as well as to the nature of the
respective reference functions.

One particularly distinguishing aspect of the proposed approaches concerns the 
complete, exhaustive representation of every possible discrete signal in the
region $\Omega$.  As such, these approaches provide the basis for systematic
studies in virtually every theoretical or applied areas involving discrete signals
or functions.  In particular, it would be interesting to revisit dynamic systems
from the perspective of the described concepts and methods, associating each
admissible signal to a respective node in the transition networks, and studying
or modeling specific dynamics by considering these networks.

The generality of the concepts and methods developed along this work paves the way
to many related further developments.  For instance, it would be interest to extend the
approach from 1D signals to higher dimensional scalar and vector fields, as well as
to other types of regions possibly including non regular borders or even disconnected
parts.  It would also be interesting to study other types of functions such as exponential,
logarithm, Fourier series, as well as several types of statistical distributions.  In addition,
the several types of obtainable transition networks can be applied as benchmark in 
approaches aimed ad characterizing classifying complex networks, as well as for
studies aimed at investigating the robustness of networks to attacks, and also from
the particularly important perspective of relating topology and dynamics in network
science.  Another interesting possibility consists in applying the developed methodologies
to the analysis of real data, such as time series, shapes and images.

Though the present work focused on undirected networks, it is possible to adapt
the proposed concepts and methods for handling directed transition networks, therefore
extending even further the possibly modeled patterns and dynamics.  this can be
done, for instance, by defining the concept of adjacency in an asymmetric manner, 
such as when one of the reference functions approaches, through incremental parameter
variations, approaches another parameterless reference function, in which case the
direction would extend from the former to the latter respectively associated nodes.
Another possibility would be to establish the directions in terms of an external
field, which could be possibly associated to a dynamical system.

Last but not least, the networks generated by the proposed methodology yield
remarkable patterns when visualized into a geometric space, presenting shapes
with diverse types of coexisting regularity, heterogeneity and symmetries. 
It has been verified that an even
wider and richer repertoire of shapes can be obtained by the suggested method
by varying the involved parameters.
For instance, symmetries of types other than bilateral can be obtained by using
reference functions containing 3 or more terms instead of the 2 terms as adopted
in most of the examples in this work.  One particularly interesting aspect of 
generating shapes in the described manner is that very few parameters are
involved while determining structures with high levels of spatial and morphologic
diversity and complexity.   Actually, the only involved parameters specifying each
of the possibly obtained shapes are $N_x$, $N_y$, the reference functions, 
$\alpha$ and $L$.   This potential for producing such flexible shapes paves
the way to several studies not only in shape and pattern generation and recognition, but also
for development biology, in the sense that the obtained structures could represent
a model of morphogenesis through gene expression control by the reference functions, while the
spatial organization of the cells would be defined in a manner similar to the
Fruchterman-Reingold method, i.e.~nodes that are connected attract one
another, while disconnected nodes tend to repel one another.  These interactions
could be associated to morphic fields (e.g.~biochemical concentrations, electric
fields, etc.) taking place during development.

\vspace{0.7cm}
\textbf{Acknowledgments.}

Luciano da F. Costa
thanks CNPq (grant no.~307085/2018-0) for sponsorship. This work has benefited from
FAPESP grant 15/22308-2 .  
\vspace{1cm}

\bibliography{mybib}
\bibliographystyle{unsrt}

\end{document}